| | |
|---|---|
| Article title: | The first order statistics of backscatter from the fractal branching vasculature |
| Authors: | Kevin J. Parker[a)] |
| | Department of Electrical and Computer Engineering, University of Rochester, Rochester, New York 14627, USA |
| Date: | 09/06/2019 |
| Running title: | The first order statistics of backscatter from the fractal branching vasculature |

[a)] Author to whom correspondence should be addressed.  Electronic mail: kevin.parker@rochester.edu

**ABSTRACT**

The issue of speckle statistics from ultrasound images of soft tissues such as the liver has a long and rich history. A number of theoretical distributions, some related to random scatterers or fades in optics and radar, have been formulated for pulse-echo interference patterns. This work proposes an alternative framework in which the dominant echoes are presumed to result from Born scattering from fluid filled vessels that permeate the tissue parenchyma. These are modeled as a branching, fractal, self-similar, multi scale collection of cylindrical scatterers governed by a power law distribution relating the number of branches at each radius. A deterministic accounting of the echo envelopes across the scales from small to large is undertaken, leading to a closed form theoretical formula for the histogram of the envelope of the echoes. The normalized histogram is found to be related to the classical Burr distribution, with the key power law parameter directly related to that of the number density of vessels vs. diameter, frequently reported in the range of 2 to 4. Examples are given from liver scans to demonstrate the applicability of the theory.

## I. INTRODUCTION

A century of research on scattering of light and sound has accumulated since the landmark papers of Rayleigh (Rayleigh, 1897; 1918a) and today, every day, an uncountable number of ultrasound scans are created from tissue backscatter. In normal soft tissues such as the liver, prostate, and thyroid, the source of the backscatter is presumed to be internal inhomogeneities which have been typically modelled as statistical inhomogeneities or spherical shapes linked to the cellular structures of the soft tissue.

For example, Chivers (1977) considered both exponential and Gaussian correlation models in his survey of tissue scattering. Waag (Waag *et al.*, 1982; Campbell and Waag, 1984b; Waag *et al.*, 1989a; Waag *et al.*, 1989b) considered Gaussian and modified Gaussian correlation models. Lizzi *et al.* (1983) established a framework for analysis of scattered ultrasound from tissues, including a model of very small scattering sites that were Poisson-distributed in space, and then later including spherical shapes that have been incorporated into quantitative ultrasound studies (Mamou and Oelze, 2013). Insana and colleagues (1990) considered three correlation functions for random media, related to fluid spheres, spherical shells, and a Gaussian behavior. These were applied to a number of phantoms and tissues (Insana and Brown, 1993). Shung and colleagues demonstrated the scattering of red blood cells which in a dispersed form exhibit classical Rayleigh scattering behavior at the frequencies commonly used in ultrasound scans of adult humans (Shung and Thieme, 1993). The importance of clustering of red blood cells was analyzed by Fontaine *et al*. (2002) and Savery and Cloutier (2001; 2005).

A recently proposed alternative hypothesis is that within soft and macroscopically isotropic tissues such as the liver, the dominant parenchymal cellular structure forms the *reference media* and the fractal branching vasculature and fluid channels form the *weak scattering struc-* *tures* within the Born approximation. The consequences of this new framework are significant, as the canonical structural shape must be cylindrical, not spherical, and the self-similar or fractal or multi-scale nature of the branching fluid channels must be considered within the ensemble average.

The second order statistics, specifically the ensemble-averaged backscatter, has been recently considered for this new hypothesis (Parker, 2019; Parker *et al.*, 2019), and the fractal branching cylindrical models were found to predict a power law $\left( f^\gamma \right)$ increase in backscatter vs. frequency for tissues such as the liver, matching the experimental results from the early leading studies (Campbell and Waag, 1984a; Zagzebski *et al.*, 1993).

However, the first order statistics of soft tissues under the new hypothesis have not yet been resolved. The distribution of the envelope of echoes from the liver and other soft tissues has been an intense subject of research over the years (Burckhardt, 1978; Bamber and Dickinson, 1980; Sleefe and Lele, 1988; Landini and Verrazzani, 1990; Wear *et al.*, 1997; Cramblitt and Parker, 1999; Kutay *et al.*, 2001; 2003; Laporte *et al.*, 2009). Based on models originating from radar and optical speckle, and on 1D convolution models, the envelope of RF echoes have been fit to Rayleigh, Gamma, Rician, Homodyne-K, and other distributions. It was hoped that some measure of tissue structure, such as the number of discrete scatterers per unit volume, could be estimated from these distribution functions, however no definitive consensus has been achieved.

With the goal of clarifying the dominant mechanisms at work in pulse-echo imaging, we depart from earlier frameworks in two significant ways: first, the scattering structures are assumed to be a multiscale set of cylindrical vessels of radius $a$ with a density distribution that follows a power law, $N(a) = N_0 / a^b$. Secondly, we derive a fully deterministic (not probabilistic) ensemble of echo amplitudes based on the dominant (maximum) signals obtained from the 3D convolution of a pulse with the isotropic ensemble of scattering cylindrical vessels. The result is a model of the envelope histogram that is deterministic and somewhat resembles – but is not equal to – the Rayleigh distribution. In this framework a key tissue parameter of the envelope distribution is the power law parameter $b$, which captures the multi-scale distribution of vessels from the few large vessels to the greater numbers of smaller vessels.

## II. THEORY

Assume a broadband pulse propagating in the $x$ direction is given by separable functions:

$$P\left(y,z,t-\frac{x}{c}\right) = G_y\left(y,\sigma_y\right)G_z\left(z,\sigma_z\right)P_x\left(t-\frac{x}{c}\right), \tag{1}$$

where $G_y\left(y,\sigma_y\right) = \exp\left[-y^2/2\sigma_y^2\right]$, i.e., Gaussian in the $y$ and $z$ directions, and where the pulse shape $P_x$ in the $x$ direction is given by:

$$P_x\left(x\right) = GH_2\left(\frac{x}{\sigma_x}\right)\exp\left(-\frac{x}{\sigma_x}\right)^2 = \mathbf{e}^{-x^2/\sigma_x^2}\left(\frac{4x^2}{\sigma_x^2}-2\right), \tag{2}$$

where $GH_2$ is a second-order Hermite polynomial for the pulse shape with a spatial scale factor of $\sigma_x$ (Poularikas, 2010; Parker, 2016), representing a broadband pulse. Its spatial Fourier transform is then:

$$^{3D}\Im\left\{P\left(x,y,z,t=0\right)\right\} = $$
$$\left(4\mathbf{e}^{-k_x^2\pi^2\sigma_x^2}k_x^2\pi^{5/2}\sigma_x^3\right)\left(\mathbf{e}^{-2k_y^2\pi^2\sigma_y^2}\sqrt{2\pi}\sigma_y\right)\left(\mathbf{e}^{-2k_z^2\pi^2\sigma_z^2}\sqrt{2\pi}\sigma_z\right), \tag{3}$$

where we use Bracewell's convention (1965b) for the form of the Fourier transform.

Using a 3D convolution model (Bamber and Dickinson, 1980; Macovski, 1983; Prince and Links, 2015), we will determine the *dominant* echoes from the pulse interacting with each generation of elements in a branching, fractal, self-similar set of vessels shown in **Figure 1**, and whose number density follows a power law behavior $N(a) = N_0 / a^b$. From these echoes, the histogram of envelopes is determined, by summing up over all the fractal branches.

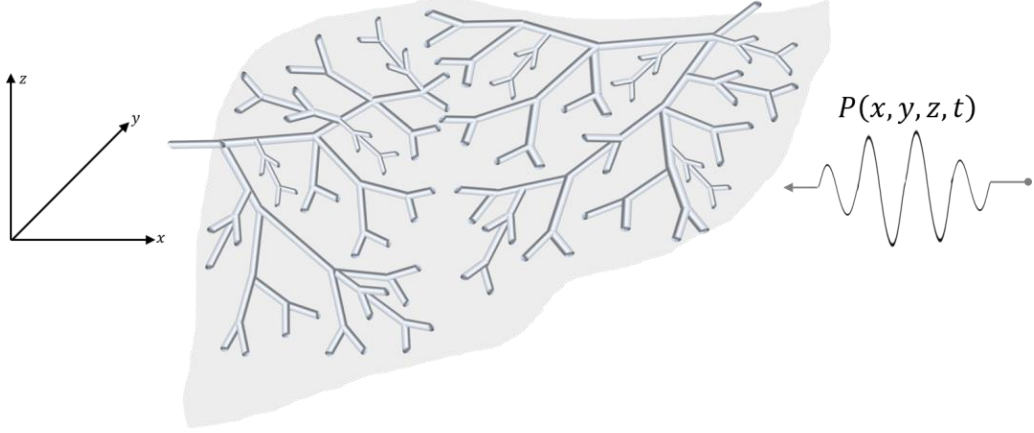

**Figure 1.** Model of 3D convolution of a pulse with the fractal branching cylindrical fluid-filled channels in a soft tissue.

Assuming an isotropic spatial and angular distribution of each generation of fractal branching structures, we need to consider a basic element across all angles of incidence with respect to the propagating wave and across all size scales, from very small micro channels of fluid to the largest arteries and veins that can exist within the organ. Specifically, we will examine a long fluid-filled cylinder of radius $a$:

$$f(r) = \begin{cases} \kappa_0 & r \leq a \\ 0 & r > a \end{cases}$$

$$F(\rho) = \frac{\kappa_0 \cdot a \cdot J_1[2\pi a \cdot \rho]}{\rho},$$

(4)

where $\kappa_0$ is the fractional variation in compressibility, assumed to be $\ll 1$ consistent with the Born formulation, $F(\rho)$ represents the Hankel transform, which is the 2D Fourier transform of a radially symmetric function, $J_1[\cdot]$ is a Bessel function of order 1, and $\rho$ is the spatial frequency. The fractional variation in compressibility, $\kappa_0$, between blood vessels and liver parenchyma has been estimated to be approximately 0.03, or a 3% difference based on published data (Parker, 2019).

Consider first one infinitely long cylinder with material property $f(r)$ – symmetric – as shown in **Figure 2(a)** and tilted at some arbitrary angle in a spherical coordinate system. It has a 3D Fourier transform that is a thin disk, shown in **Figure 2(b)** (delta function $k_z$ but shown with finite thickness to make the graphic easier to draw and visualize). A particular radius of spatial frequency equal to $q_0$ is shown for reference. The shape of the 3D spatial Fourier transform of the pulse is shown in **Figure 2(c)**.

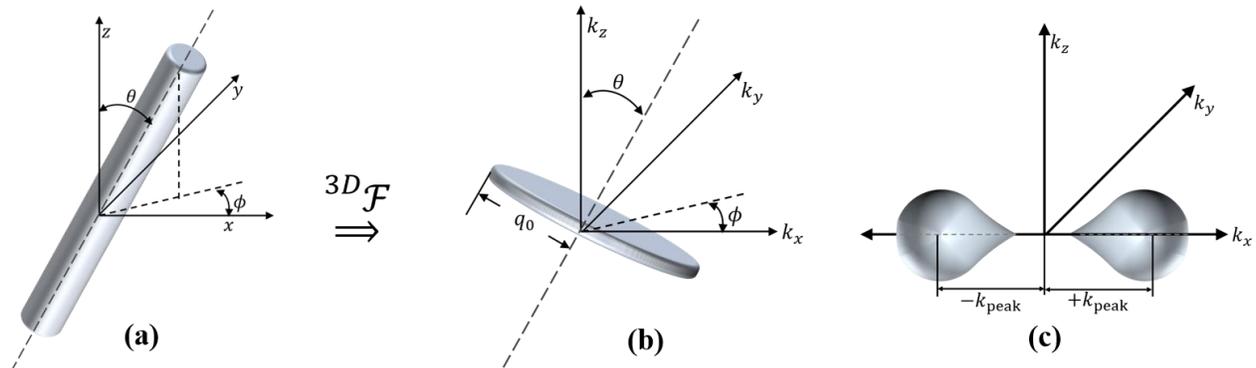

**Figure 2.** A cylindrical function (a) and its Hankel transform represented in 3D Fourier transform space (b). Rotations around spherical coordinates similarly rotates the corresponding transform. The transform of a propagating pulse is shown in (c). The maximum product of (b) and (c) arrives when the angle $\theta$ approaches zero.

One can see from **Figure 2** that, the convolution of the pulse with a cylinder of radius $a$ is dominated by the case where the cylinder is perpendicular to the direction of the forward propagating pulse, the $x$ – axis in our case. For all other orientations, the delta function of the cylindrical transform shown in **Figure 2(b)** is misaligned with respect to the transform of the pulse shown in **Figure 2(c)**. Thus, assuming an optimal alignment, the 3D convolution result is given by the product of the transforms:

$$^{3D}\Im\left\{echo\left(x,y,z,t=0\right)\right\}=\Im^{3D}\left\{p\left(x,y,z\right)\right\}\bullet\left(k_x\right)^2\Im^{3D}\left\{cylinder\left(x,y,z\right)\right\}=$$

$$\left(\left(4\mathbf{e}^{-k_x^2\pi^2\sigma_x^2}k_x^2\pi^{5/2}\sigma_x^3\right)\left(\mathbf{e}^{-2k_y^2\pi^2\sigma_y^2}\sqrt{2\pi}\,\sigma_y\right)\left(\mathbf{e}^{-2k_z^2\pi^2\sigma_z^2}\sqrt{2\pi}\,\sigma_z\right)\right)\bullet \tag{5}$$

$$\left(k_x\right)^2\kappa_0\left(a\left[\frac{1}{\sqrt{k_x^2+k_y^2}}\right]\bullet\left(J_1\left[2a\pi\sqrt{k_x^2+k_y^2}\right]\right)\right)\delta\left[k_z\right],$$

where $t=0$ is assigned to the central peak of the echo, and the $\left(k_x\right)^2$ term pre-multiplying the cylinder transform stems from the Laplacian spatial derivative in the Born scattering formulation (Rayleigh, 1918a; Morse and Ingard, 1987) and in the 3D convolution model (Gore and Leeman, 1977; Bamber and Dickinson, 1980).

By Parseval's theorem, the integral of the square of the transform equals the integral of the square of the echo, and after integration over the delta function in $k_z$:

$$\int\left(\text{echo}\left(t\right)\right)^2dt=\sigma_z^2\kappa_0^2\int\limits_{kx=-\infty}^{\infty}\int\limits_{ky=-\infty}^{\infty}\left(8\mathbf{e}^{-\pi^2\left(k_x^2\sigma_x^2+2k_y^2\sigma_y^2\right)}k_x^2\pi^{7/2}\sigma_x^3\sigma_y\right)^2\bullet$$

$$\left(k_x^2\right)^2\left(a\left[\frac{1}{\sqrt{k_x^2+k_y^2}}\right]\bullet\left(J_1\left[2a\pi\sqrt{k_x^2+k_y^2}\right]\right)\right)^2dk_xdk_y. \tag{6}$$

The square root of this gives the root mean square (RMS) amplitude of the echo, as a function of the radius $a$, shown for the practical span of $0<a/\sigma_x<10$ in **Figure 3**. We will associate the RMS amplitude from each echo with a proportionally higher maximum value of the envelope, as a function of cylinder radius $a$. This single value mapping is justified in the **Appendix.**

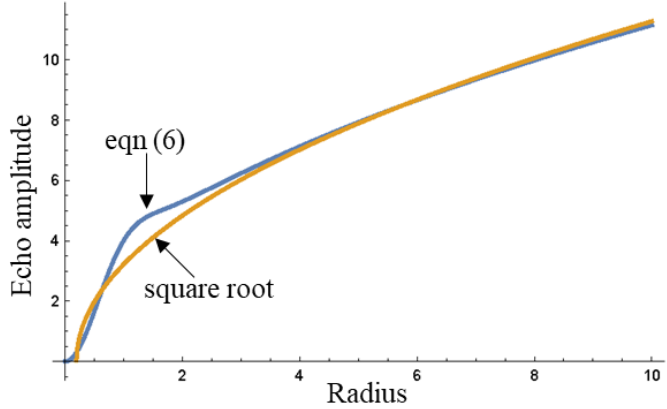

**Figure 3.** The RMS echo amplitude (vertical axis, arbitrary units) vs. cylinder radius normalized by $\sigma_x$ from eqn (6), and square root approximation, assuming a $GH_2$ pulse.

Also shown in **Figure 3** is an approximation which will be useful for deriving a closed form solution, the approximation is of the form $A[a] = A_0 \sqrt{a - a_{\min}}$, justified by the nearly linear increase in the energy term above some minimum threshold, and the asymptotic modulus of $J_1(ak)$ which is proportional to $\sqrt{2/(\pi ak)}$ (Abramowitz and Stegun, 1964) as $ak$ becomes large. Of course the exact shape is dependent on the particular pulse shape's spectrum, for example if instead of a GH2 we use a $k_x \cdot \mathrm{sech}[k_x]$ bandpass for $\Im\{P_x(x)\}$, which has an exponential asymptotic tail instead of a Gaussian, then the result is shown in **Figure 4**:

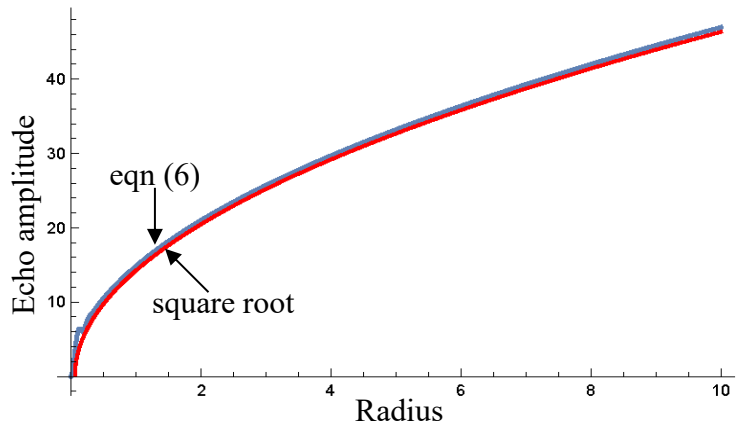

**Figure 4.** The RMS echo amplitude for a second bandpass pulse shape with exponential tails vs. normalized radius of a scattering cylinder, and a square root approximation.

So as a general approximation, we apply the relation $A[a] = A_0 \sqrt{a - a_{\min}}$ for $a > a_{\min}$. The parameter $a_{\min}$ depends on a number of factors, including the dynamic range selected (for example, 45 dB) and the Rayleigh scattering (long wavelength, small $a$) behavior of the cylinder interacting with the particular pulse transmit signal, along with the noise floor and quantization floor of the receiver.

Now, applying the general theory of transformed distributions (see **Appendix**), we have within the ensemble the number density of vessels at different radii given by $N[a] = N_0 / a^b$, and this will be transformed into the distribution of amplitudes, $A(a)$. The general rule is:

$$N[A] = \frac{1}{dA/da} N[a]. \tag{7}$$

In our case, the derivative $dA/da = \left[ (1/2) A_0 \right] / \sqrt{a - a_{\min}}$, and the inverse function is $a[A] = (A/A_0)^2 + a_{\min}$. Thus, substituting these into eqn (7) the distribution $N[A]$ is:

$$N[A] = \frac{2 N_0 A}{A_0^2 \left[ (A/A_0)^2 + a_{\min} \right]^b} . \tag{8}$$

So, for example, if $b = 2$ and $A_0$ and $N_0$ are unity, then $N[A] = 2A / \left( A^2 + a_{\min} \right)^2$, and this is plotted in **Figure 5** along with variations in parameters.

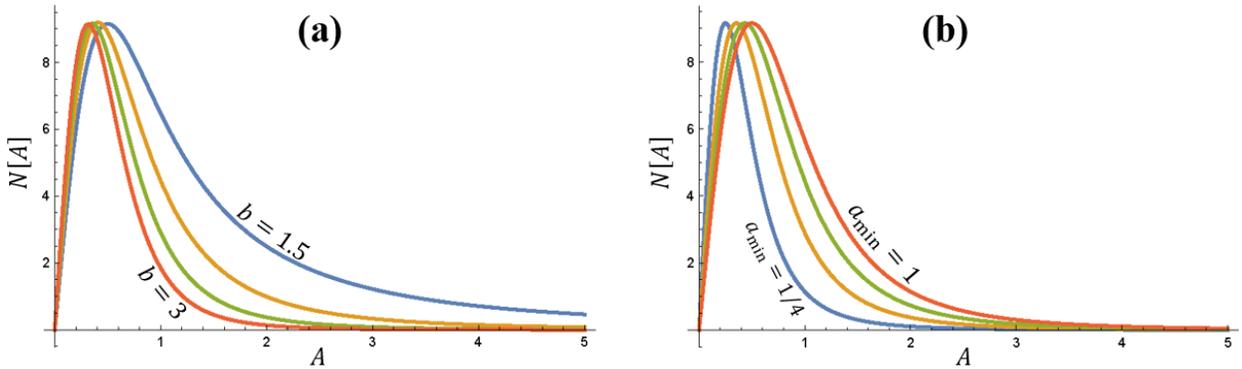

**Figure 5.** The proposed histogram function of envelope amplitudes $A$, having the form $A/\left(A^2 + a_{min}\right)^b$. In (a) are normalized functions where $a_{min} = 1/2$ and the power law parameter $b$ is 3, 2.5, 2, and 1.5. In (b) are normalized functions where the power law parameter is fixed at 2.5 however $a_{min}$ is varied as 1/4, 1/2, 3/4, and 1. Vertical axis: counts (arbitrary units); horizontal axis: envelope amplitude (arbitrary units).

This provides a four-parameter fit $\left\{N_0, A_0, a_{min}, b\right\}$ to a histogram taken from a reasonably sized region of interest (ROI) within a vascularized tissue or organ, assuming an isotropic and spatially uniform distribution across the ROI.

However, of these four parameters, $N_0$, $A_0$, and $a_{min}$ are influenced by system parameters such as amplifier gain and the size of the ROI. To simplify the analysis, one can normalize by the integral of the distribution $\int N\left[A\right] dA = N_0 / \left[\left(b-1\right)\left(a_{min}\right)^{b-1}\right]$ to form a proper probability density function (PDF), which integrates to unity:

$$N_n\left[A\right] = \frac{2A\left(a_{min}\right)^{b-1}\left(b-1\right)}{A_0^2\left[\left(\frac{A}{A_0}\right)^2 + a_{min}\right]^b}. \qquad (9)$$

Furthermore, by substituting $\lambda = A_0\sqrt{a_{min}}$, we find this reduces to a two-parameter distribution:

$$N_n\left[A\right] = \frac{2A\left(b-1\right)}{\lambda^2\left[\left(\frac{A}{\lambda}\right)^2 + 1\right]^b}, \qquad (10)$$

which is a Burr Type XII distribution with $c = 2$ (Burr, 1942; Rodriguez, 1977). Thus, the normalized distribution offers a simplification to a two-parameter distribution with analytic expressions for probability density function, cumulative distribution function, and moments (Rodriguez, 1977).

Using the notations from Bracewell (1965a), we calculate by integration the centroid of the distribution and find the expected value of the normalized histogram:

$$\left\langle A \right\rangle = \frac{\lambda \left(-1+b\right)\sqrt{\pi}\,\Gamma\left[-\dfrac{3}{2}+b\right]}{2\Gamma\left[b\right]} \text{ for } b > \frac{3}{2}, \tag{11}$$

where $\Gamma\left[\cdot\right]$ is the gamma function (Abramowitz and Stegun, 1964). Similarly, the mean square abscissa can found by integration:

$$\left\langle A^2 \right\rangle = \frac{\lambda^2}{b-2} \quad \text{for} \quad b > 2. \tag{12}$$

In a similar way, the variance can be calculated, and then a measure of signal-to-noise (ratio, SNR) defined by the mean value over the standard deviation is:

$$\text{SNR} = \frac{\left\langle A \right\rangle}{\sigma_A} = \frac{\Gamma\left[-\dfrac{3}{2}+b\right]}{\sqrt{4\left(b-2\right)\Gamma\left(b-2\right)^2\big/\pi-\Gamma\left(-\dfrac{3}{2}+b\right)}}. \tag{13}$$

Generally, in the range of interest ( $2 < b < 4$ ), we find that the SNR is less than 1.5, which is below the theoretical 1.91 SNR found for fully-developed speckle under a Rayleigh probability density function (Burckhardt, 1978; Tuthill *et al.*, 1988; Thijssen, 2003). However, it must be kept in mind that our underlying model of cylinders vs. radius has a power law tail as the radius *a* goes to infinity, creating a long asymptotic tail of high echo amplitudes. In practice, each organ has a finite upper limit to the largest artery or vein, and this truncates the upper tail of the distribution, leading to a smaller standard deviation and a larger SNR than would be suggested by formulas.

## III. METHODS

In the following examples, conventional B-scans were obtained using a Verasonics scanner with a 5 MHz ATL linear array transducer (V1, Verasonics, Inc., Kirkland, WA, USA). A scan from a healthy adult was imaged under the requirements of informed consent and the University of Rochester Institutional Review Board. Rat experiments were reviewed and approved by the Institutional Animal Care and Use Committee of Pfizer, Inc. Groton Connecticut, where the ultrasound scan was acquired using a Vevo 2100 (VisualSonics, Toronto, CA) scanner and a 20 MHz center frequency transducer (data provided courtesy of Terry Swanson). Parameter estimation was performed using MATLAB (The Mathworks, Inc., Natick, MA, USA) nonlinear least squares minimization of error, for two-parameter fits of the Burr distribution to the data.

## IV. RESULTS

An ultrasound B-scan of a normal rat liver is shown in **Figure 6 (left)**, with a region of interest denoted within the liver parenchyma where the pattern of echoes demonstrates a speckle pattern. This region is distal to the transmit focus, which is located at 11 mm depth. A close-up view is shown in **Figure 6 (right)**.

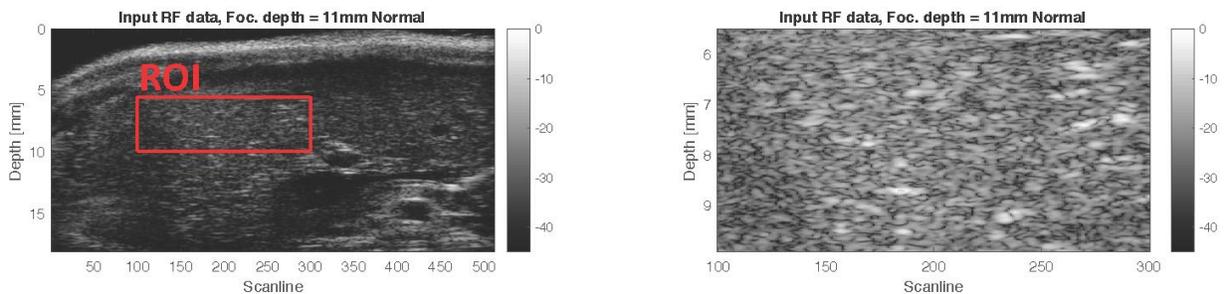

**Figure 6.** Left: ultrasound B-scan image from a 20 MHz scan of a normal rat liver, focused transmit at 11 mm depth. Right: zoom view of speckle region ROI selected for analysis.

The envelope of the beam-formed RF is distributed as shown in the histogram of **Figure 7**, along with a theoretical Burr distribution with a power law parameter of $b = 3.4$, and an $R^2$ of 0.999. The match of this deterministic theory to the data is reasonable.

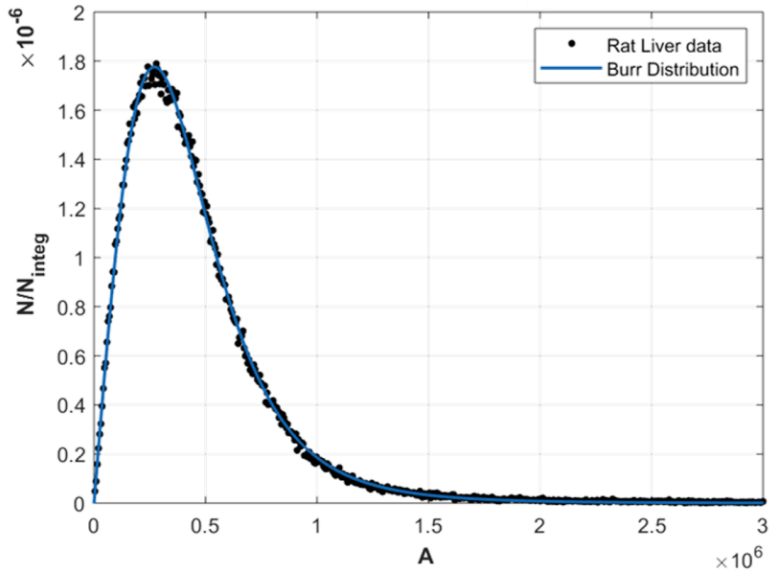

**Figure 7.** Histogram of echo amplitudes from the normal rat liver. Vertical axis = counts; horizontal axis = amplitudes of echo envelopes (arbitrary units). The dashed line indicates theoretical fit to the derived eqn (10), with the power law parameter of $b = 3.4$ and $\lambda = 6.5 \times 10^5$.

The normal human liver is shown in **Figure 8**. This utilized five compounded plane wave transmit pulses with dynamic receive.

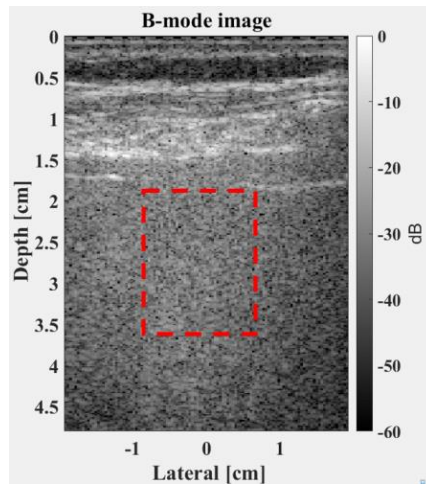

**Figure 8.** B-scan image of healthy liver tissue in a human. A region of interest is selected (dotted lines) for analysis.

The histogram of echo amplitudes is shown in **Figure 9** along with the Burr distribution fit with a power law parameter of $b = 2.9$ and an $R^2$ of 0.995.

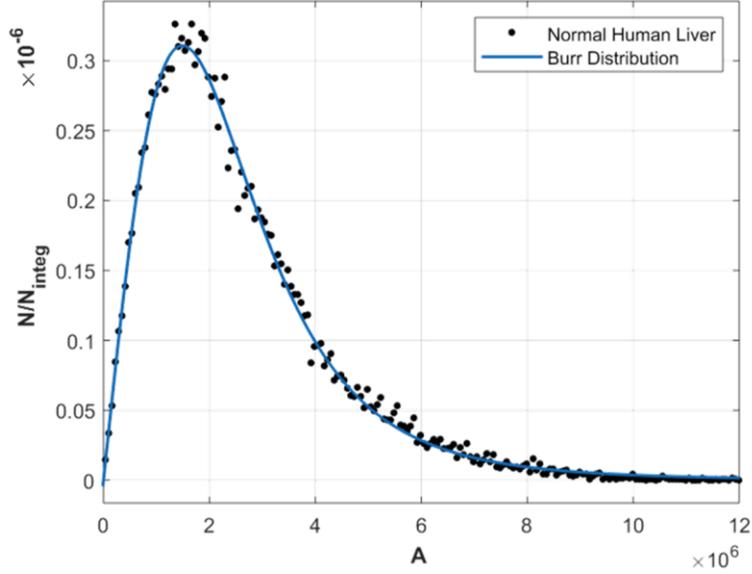

**Figure 9.** Data from a 5 MHz probe on a normal human liver with $b = 2.9$ and $\lambda = 3.3 \times 10^6$. Vertical axis: counts; horizontal axis: amplitudes of echo envelopes (arbitrary units).

To further examine the asymptotic characteristics of the histogram, we integrate the normalized histogram from the rat liver data to form the cumulative distribution function (CDF) which is important in a number of statistical tests such as the Kolmogorov-Smirnov test (Feller, 1948). This is shown in **Figure 10** using the parameters found in **Figure 7** ( $b = 3.4$ and $\lambda = 6.5 \times 10^5$ ) against the theoretical CDF for the Burr distribution, given by:

$$\text{CDF}\left(A\right) = 1 - \left[ 1 + \left( \frac{A}{\lambda} \right)^2 \right]^{-(b-1)}.$$

(14)

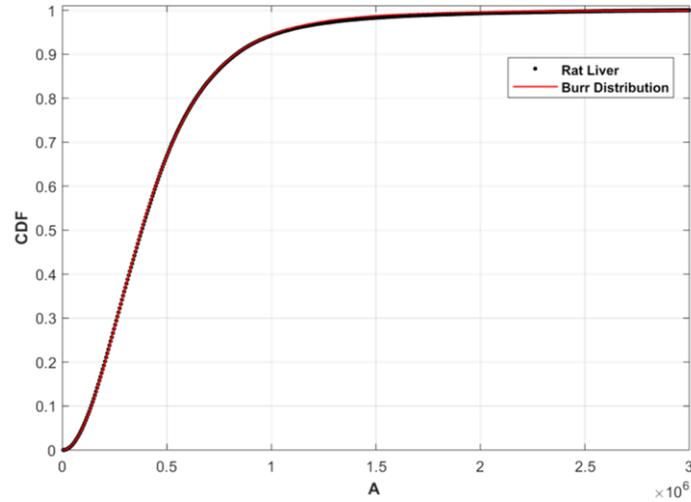

**Figure 10.** The comparison of the cumulative distribution function (CDF) of the rat liver echo amplitudes and the theoretical Burr distribution CDF.

The two functions (sampled histogram counts and theoretical function) are nearly indistinguishable on this scale and resolution.

Finally, because of the historic interest in the Rayleigh distribution (Burckhardt, 1978), the PDF for the rat liver is shown with the minimum mean squared error curve fit, shown in **Figure 11**, where the Rayleigh PDF is given by:

$$R(A) = \frac{A}{S^2} \mathbf{e}^{\left(-\frac{A^2}{2S^2}\right)}. \tag{15}$$

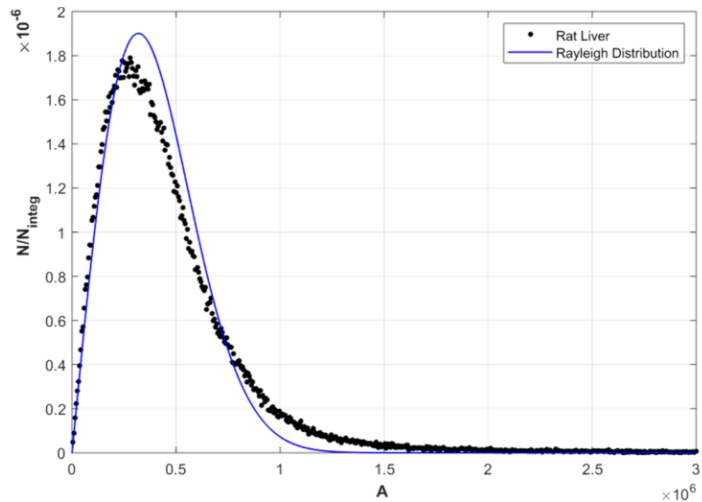

The overall shape of the Rayleigh distribution appears more compact and less suitable for the measured amplitude data.

## V. DISCUSSION AND CONCLUSION

This framework postulates that the histogram of speckle amplitudes is governed by a simple equation eqn (8) governed by four parameters: a power law $b$, a minimum cylinder size $a_{min}$ above which the echo amplitude rises as a square root of radius, a vessel density $N_0$, and a scale amplitude $A_0$ depending on system factors such as amplifier gain. Furthermore, upon normalization of the histogram by its integral, the resulting PDF reverts to a Burr distribution first described in 1942 without any consideration of waves or scattering. The formulation is a major departure from the treatments over the last 100 years in that the accounting is deterministic and focused on the maximum or predominant result in each step. Effectively, this means that each local maximum of the echoes is modeled as resulting from one dominant cylinder aligned perpendicular to the propagation direction of the pulse. Then, assuming a spatial ensemble large enough to encompass all radii according to the power law, the form of eqn (8) is derived as the histogram of speckle envelopes within soft vascularized tissue. This formulation is limited by the major assumptions which simplify the derivations:

- The echoes are assumed to be produced by an isotropic ensemble of long cylindrical fluid-filled Born scatterers, dominated by those cylinders that are aligned perpendicular to the propagating pulse.

- The echoes from each generation of cylinders of radius $a$ are mapped to an envelope amplitude of $A$ by a square root function.

- The local maxima of envelopes dominate the histogram of sampled echoes.

Each of these assumptions are linked to a theoretical formula with examples, however collectively they simplify the accounting of the overall chain of echo formation into a final histogram. Consequently, the issues of complex interference of phasors are not included in the analysis. In this sense our analysis returns to the earliest formulations of weak scattering from cylinders (Rayleigh, 1918b; Albini and Nagelberg, 1962).

An important question which must be resolved by further studies of ensembles of human and animal soft tissue echoes is the following: will one of the theoretical distributions (Burr, Rayleigh, Rician, K, Homodyne-K, Gamma, or others) described in the literature be consistently the best fit, with lower mean squared errors, than all the others? A related question involves the sensitivity of the parameters to abnormalities that develop within the tissues and that change the scattering characteristics. Mathematically, we can say that the Burr distribution derived herein has an asymptotic behavior for high amplitudes that includes a power law proportional to $A^{-2b}$. In comparison, the other distributions derived in the classical literature have asymptotic behaviors related to Gaussian (Rayleigh), Gaussian times modified Bessel $I_0$ function (Rician), modified Bessel K functions (K distribution), or other transcendental functions (Gamma). The fractal branching vasculature hypothesis for scattering is naturally associated with a power law behavior describing the number of vessels of different diameter, and this power law $b$ propagates through the derivations of the scattering functions and the final accounting of histogram amplitudes. Thus, a power law asymptotic behavior in amplitudes is plausible.

Finally, the important task of diagnosing abnormalities is left for further research. Presumably a disruption of the normal vascular structure in aggressive cancers would change the power law $b$ and the vessel number density $N_0$, but the necessary clinical studies of these and other common pathologies are beyond the scope of this paper.

Further high resolution studies of the vasculature in 3D for different organs, along with corresponding high resolution B-scans of the same organs, would be helpful for testing and refining this theoretical framework of speckle amplitudes from soft vascularized tissues.

## ACKNOWLEDGEMENTS

This work was supported by National Institutes of Health grant number R21EB02590. The author is grateful to PhD students Sedigheh Poul, Juvenal Ormachea, and Jihye Baek for providing envelope data and curve fits, and to Terry Swanson of Pfizer Inc., Groton CT, for providing the superb B-scan of the rat liver.

## APPENDIX: MAPPING AND TRANSFORMATION OF FUNCTIONS

Herein we consider how the size distribution of cylindrical elements determines a distribution of echoes, in order to derive a histogram of echo amplitudes. A lucid explanation of the mapping of distributions is given in Chapter 5 of Papoulis (1987). This is explained in terms of probability density functions, where a new variable is defined by $y = g(x)$, and $x$ is a random variable with known probability density function $f(x)$. However, the results hold for all well-behaved analytic functions where the area under the curve within any small interval is preserved by the mapping. We also assume a single valued mapping of $y = g(x)$ and its inverse $x = g^{-1}(y)$ for simplicity. In that case, by simply equating equal areas under some region of $x_0$ to $x_0 + \Delta x$, and mapping that to the $y$ variable, the transformation rule is:

$$f_y(y) = \frac{f_x(x)}{|dg(x)/dx|}, \tag{16}$$

where on the right side the inverse function $x = g^{-1}(y)$ is then used to eliminate $x$ as a variable and produce an equation in terms of $y$. The concept is illustrated in **Figure 12**.

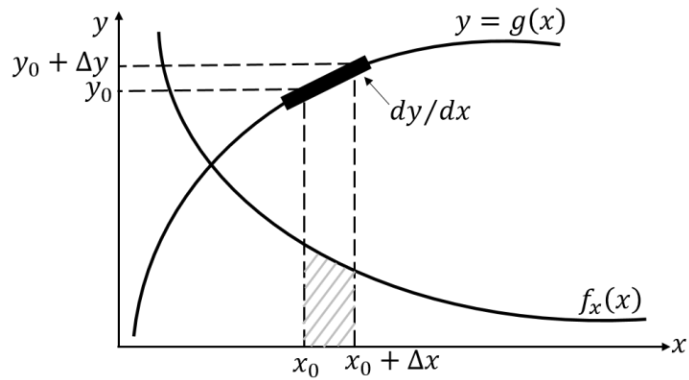

**Figure 12.** Schematic illustration following Papoulis showing the distribution of one continuous variable $x$ with a distribution $f_x(x)$ which is mapped to a new distribution where $y = g(x)$, and retaining the area under the curve.

An important feature of this transformation for echo envelope functions emerges as a dramatic consequence, for anywhere near to a local maxima where the first derivative of the envelope approaches zero, the denominator term in eqn (16) approaches zero and the transformation produces a singularity around that point. Thus, as we apply this to echo envelopes where $x$ is the echo space or time variable and $y$ is the echo envelope, each local maximum creates a large spike in the histogram of the envelope.

As an example, let $y = A_1(1 - x^2)$ for $-1 < x < 1$. By symmetry, we only need to consider $0 < x < 1$ and assume uniform likelihood of sampling within this interval so $f_x(x) = 1$. Also,

$|dy/dx| = A_1 \cdot 2x$ and $x = g^{-1}(y) = \sqrt{1 - (y/A_1)}$. Finally, from eqn (16),

$f(y) = 1 / \left[ 2A_1\sqrt{1 - (y/A_1)} \right]$ for $0 < y < A_1$, which has a singularity as $y \to 1$ (the value of the local maximum of $y = f(x)$). This is illustrated in **Figure 13** below for $A_1 = 1$.

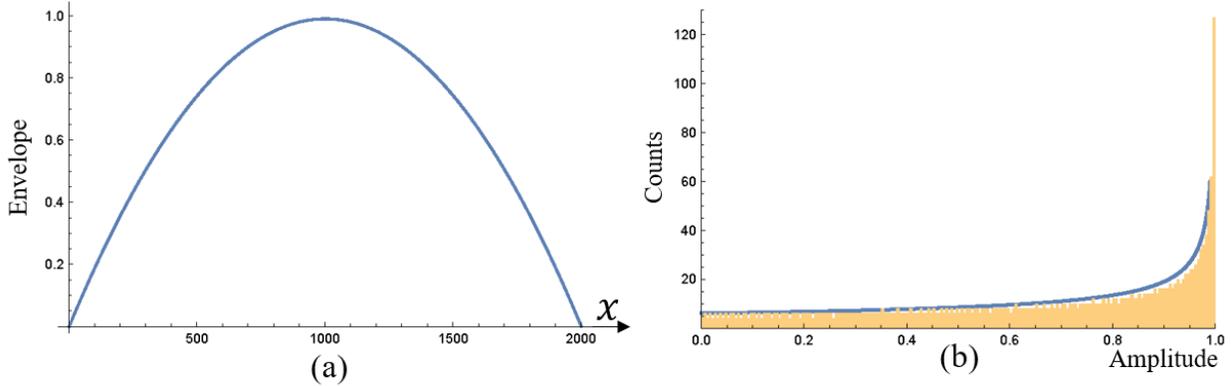

(a)  (b)

**Figure 13.** Example of sampling an envelope function (a) to produce a resulting histogram (b) and theory from eqn (16) (solid line). In (a), the vertical axis is amplitude, maximum value normalized to 1; the horizontal axis is time or distance, arbitrary units. In (b), the vertical axis is sampled counts in a histogram; the horizontal axis is echo amplitude. Note the spike near amplitude=1, the local maximum of the envelope.

Another example of a modified Gaussian is shown in **Figure 14**, where $y = \mathbf{e}^{-x^2/2}(1 - x^2)$, producing a similar result deriving from the singularity produced by the local maximum of the envelope function.

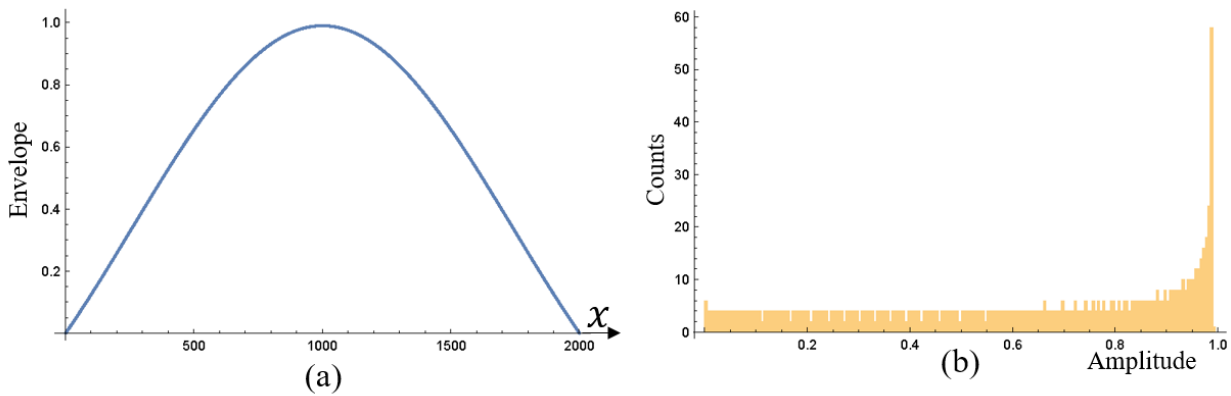

(a)  (b)

**Figure 14.** Second example using a modified Gaussian envelope in (a) and the resulting histogram in (b), demonstrating the dominant value or singularity mapped from the local maximum. In (a), the vertical axis is amplitude, maximum value normalized to 1; the horizontal axis is time or distance, arbitrary units. In (b), the vertical axis is sampled counts in a histogram; the horizontal axis is echo amplitude. Note the spike near amplitude=1, the local maximum of the envelope.

Thus, we conclude that the histogram of envelopes contains a dominant contribution formed by the local maximum, where the first derivative approaches zero. This dominant contribution from the local maximum of the echo is also assumed to be proportional to the RMS echo amplitude formed by convolution, eqn (6), and then is used in the superposition or summation over all cylinder sizes in our formulation.